\documentclass[10pt, journal]{IEEEtran}
\usepackage{todonotes}
\usepackage{cleveref}
\usepackage[hidelinks]{hyperref}
\usepackage{pdflscape}
\usepackage{afterpage}
\usepackage{booktabs}
\usepackage{listings}
\usepackage{capt-of}
\newcommand{\stateoftheart}{state-of-the-art}
\newcommand{\blockchain}{blockchain}
\newcommand{\property}[1]{\textit{#1}}
\newcommand{\selfsovereign}{SSI}

\newcommand{\refappendix}[1]{\hyperref[#1]{Appendix~\ref*{#1}}}
\begin{document}
\title{Self-Sovereign Identity Solutions: The Necessity of Blockchain Technology.}

\author{\IEEEauthorblockN{Dirk van Bokkem\IEEEauthorrefmark{1},
Rico Hageman\IEEEauthorrefmark{2}, Gijs Koning\IEEEauthorrefmark{3},
Luat Nguyen\IEEEauthorrefmark{4} and Naqib Zarin \IEEEauthorrefmark{5}}
\IEEEauthorblockA{\\Computer Science and Engineering\\
Delft University of Technology\\
Email: \IEEEauthorrefmark{1}d.vanbokkem@student.tudelft.nl,
\IEEEauthorrefmark{2}r.hageman-1@student.tudelft.nl,
\IEEEauthorrefmark{3}G.Koning@student.tudelft.nl,
\IEEEauthorrefmark{4}t.l.nguyen@student.tudelft.nl,
\IEEEauthorrefmark{5}n.zarin@student.tudelft.nl}
}

\maketitle

\begin{abstract}
With the advancement of technology and subsequently the age of digital information, online trustworthy identification has become increasingly more important. With respect to the various data breaches and privacy regulations, the current identity solutions are not fully optimized. In this paper, we will take a look at several Self-Sovereign Identity solutions which are already available. Some of them are built upon \blockchain\ technology as this already provides decentralized persistent data and consensus. We will explore the emerging landscape of Self-Sovereign Identity solutions and dissect their implementations under multiple aspect criteria to determine the necessity of \blockchain\ technology in this field. We conclude that blockchain technology is not explicitly required for a Self-Sovereign Identity solution but it is a good foundation to build up on, due to various technical advantages that the blockchain has to offer.  

\end{abstract}


\section{Introduction}
\label{sec:intro}

\IEEEPARstart{A}{s} the internet grows rapidly, saving and accessing sensitive information of users is becoming a serious problem. The internet lacks a layer of identity protocol and this shifts the responsibility for identification and verification to service providers. Over time, the providers have become the centralized authorities and acted as the issuers and authenticators of digital identity. This is not only highly inefficient due to duplication of the information among them, it also prevents users to gain insight in and control over their personal digital identity. Since most of these identity management systems rely on centralized databases, it poses a threat to the user when compromised \cite{survey}. 

New identity management schemes , solving the above-mentioned problems by utilizing the same digital identity on different sites, were proposed. Some currently available examples are Facebook Login and Google Login. One does not need a username and password anymore to sign up for a particular platform, reducing the information duplication problem. Also, since the federated instances like Facebook and Google are trusted to have a secure digital identity policy, it seems that the authorization problem is also tackled. However, this means that users have to rely on the federated instances and trust these instances which makes them powerful. This way the users still have no control over their digital identity since they do not know what data is exactly collected and what it is used for. A recent example is the Cambridge Analytica scandal where Facebook gave unfettered and unauthorized access to personally identifiable information (PII) of more than 87 million Facebook users without their consent to the data firm Cambridge Analytica \cite{8436400}.

Self-Sovereign Identity (SSI) is a more recent solution to this problem and aims to give users control over their own digital identity. SSI removes the need for a central trusted authority \cite{criteria-11}. Users can store their identity data on local devices and provide the required information to those who need it for validation purposes. Bitcoin has played an important role in the SSI evolution because of its underpinning Distributed Ledger Technology (DLT) \cite{8425607}. DLT seems to be promising for SSI since it does not require a central authority to validate transactions. Due to its append-only nature, all the published attributes are persistently stored. It is also transparent because the decentralized network is able to reach consensus in the network \cite{criteria-11}. Although \blockchain\ seems promising there are some limitations with respect to SSI. For example, when users loose their private/public key pair, they are forced to start the identity proofing process from scratch to reestablish their digital identity \cite{negativeBlockchain}.

Current literature mainly focuses on individual SSI solutions. This paper contributes to the literature by comparing currently available blockchain-based and other \stateoftheart\ solutions.
The main question we hope to answer after this study is: \textit{Is blockchain-technology a necessity in building Self-Sovereign Identity solutions?}
In order to answer this question the following questions have to be answered first:
\begin{enumerate}
    \item \textit{How can Self-Sovereign Identity solutions be compared?}
    \item \textit{How do currently available solutions score based on the comparison criteria?}
\end{enumerate}

We start with introducing the current field of available evaluation criteria and comparative studies in \autoref{sec:related_work}. We follow up with the systematic search performed to gather all the available implementations, comparative studies and other sources on SSI in \autoref{sec:research_methodology}. After this, in \autoref{sec:selection_of_evaluation_criteria}, we elaborate on the evaluation criteria selected for this paper followed by the evaluation of the found implementations in \autoref{sec:evaluation}. An overview will be provided by presenting two tables containing all the properties met by all the blockchain-based and other implementations. Finally, we will answer the main research question in \autoref{sec:conclusion}. 

\section{Preliminaries}
\label{sec:preliminaries}
To get a better understanding of this paper some definitions and abbreviations are explained in more detail.
 
\textit{Identity system}: Electric information associated with an individual in a particular identity system is called a digital identity. Identity systems can be used for authentication and authorization of these identities \cite{digitalID}.

\textit{Federated instance}: or a federated identity in information technology is the process of linking a person's electronic identity and attributes, stored across multiple distinct identity management systems.

\textit{Personally identifiable information (PII)}: Any information that could potentially identify a person. Examples include full name, social security number and e-mail address. 

\textit{Distributed Ledger Technology (DLT)}: A distributed ledger is a consensus of replicated, shared, and synchronized digital data geographically spread across multiple sites, countries, or institutions \cite{8425607}.

\textit{Know-your-customer (KYC)}: a process to verify the identity of a user and calculate the potential risks of illegal intentions to a business.

\textit{Anti-money laundering (AML)}: refers to procedures, laws and regulations designed to stop the practice of generating income through illegal activities.

\textit{Decentralized Identifier (DID)}: Independent self-controlled identifier used to resolve to a DID document containing all the information required to interact with the identity.

\textit{Out-of-band}: Communicating data through another medium than the one at hand. For example; authentication through the web, with an extra out-of-band authentication via telephone.

\textit{Smart contract}: Code on the \blockchain\ that stores rules of an agreement, automatically verifies the fulfilment of this agreement and eventually executes the agreed terms.

\section{Related work}
\label{sec:related_work}
In this section we discuss some of the available evaluation criteria for \selfsovereign\ solutions and how they relate to each other. Next to that, we provide a summary of the comparative studies on the subject and which criteria are used in their comparison.

In \textit{The Laws of Identity} by K. Cameron \cite{criteria-7}, seven laws are described that explain the successes and failures of digital identity systems. K. Cameron states that these laws are necessary to avoid any side-effects. The laws and their explanation are extensive and explain the requirements of \selfsovereign\ solutions in detail. However, some of these laws can be more distinctive. The first law for example; \textit{User control and consent} could be split into \textit{Control} and \textit{Consent}. Some implementations may satisfy one of these properties, but not the other.

C. Allen used these seven laws to compose a list of ten principles that ought to be considered when implementing an \selfsovereign\ solution \cite{criteria-10}. These ten principles focus on the user control within \selfsovereign\ solutions. C. Allen provides the distinction of the seven laws we were looking for.

In \textit{Deployment of a Blockchain-Based Self-Sovereign Identity} by Q. Stokkink and J. Pouwelse \cite{criteria-11}, these ten principles of C. Allen are used in assessing the digital identity solution presented in their paper. Q. Stokkink and J. Pouwelse add an extra property to this list, which requires claims to be provable.

X. Zhu and Y. Badr explored the possibilities of current available authentication solutions for the internet of things \cite{zhu_badr_2018}. Although attestations about an identity are not required for the internet of things, several aspects do overlay like scalability, as a large population must be capable to use the system, and interoperability to prevent reliance on a single provider. It briefly touches upon multiple implementations used in this survey but does not conclude anything.

In \textit{A First Look at Identity Management Schemes on the Blockchain} \cite{8425607} a comparison between Sovrin, uPort and ShoCard with respect to the seven laws of identity of K. Cameron is made. Sovrin and uPort are \selfsovereign\ solutions whereas ShoCard is called a decentralized trusted identity. With ShoCard, identity proofing of users based on existing trusted credentials is stored on a blockchain. It is concluded that distributed ledger technology is not a silver bullet and especially the usability aspect has to improve.

A. G. Nabi provides in \textit{Comparative Study on Identity Management Methods Using Blockchain} \cite{Nabi_2017} an  overview of 23 identity solutions and their inner workings together with their advantages and disadvantages. Although seemingly it is touching the subject of comparing the solutions, it does not clearly conclude which solution is better than others.

\section{Research methodology}
\label{sec:research_methodology}
To answer the research question, a complete global view of the current \stateoftheart \ solutions is required. In this section the systematic search, constructed in three steps, for published solutions is explained. The initial search for papers about SSI implementations consisted of structured queries, as available in \refappendix{researchquery}, on both Scopus and IEEexplore. This resulted in 170 unique papers after the removal of 20 duplicates.

Secondly, these papers were categorized in \textit{implementations}, \textit{evaluation criteria}, \textit{comparative studies} and \textit{others}. The found implementations were directly included in the scope of this survey. Comparative studies were scanned to collect the different implementations, which were investigated further. This resulted in 32 different implementations.

Finally all the found identity solutions were read completely to extract the specifications according to the used evaluation criteria. In this process 20 of the found solutions were discarded as they turned out to not be SSI solutions.

\section{Selection of evaluation criteria}
\label{sec:selection_of_evaluation_criteria}
Evaluation criteria are necessary to be able to discuss and reason about the need of \blockchain -technology in the field of \selfsovereign . The criteria used in this paper are formulated as properties, which will be presented in this section.

The main ideas behind the seven laws of K. Cameron \cite{criteria-7} are incorporated in the properties by C. Allen \cite{criteria-10}, but formulated in a more comprehensive way. By splitting these seven laws into more distinct properties, the \selfsovereign \ solutions can be assessed more in-depth. Q. Stokkink and J. Pouwelse make a valid point by arguing that claims of a user do not mean anything if you cannot prove them \cite{criteria-11}. Each of these papers builds upon the previous one and the found eleven properties seem to be a complete list of features SSI solutions ought to have.
 
Therefore, the ten properties described by C. Allen, accompanied with the property of claims being provable, will form the basis of this comparative study. Below these eleven properties are quoted from the paper by Q. Stokkink and J. Pouwelse. For consistency, each property is complemented with our interpretation.
\begin{enumerate}
    \item \textbf{Existence.} \textit{Users must have an independent existence.} \\
    An SSI should be based on an identity in the real world and cannot exist exclusively in the digital world. At any time, a person should be able to independently create a digital identity, without the intervention of a third party.
    \item \textbf{Control.} \textit{Users must control their identities.} \\
    Users have the full authority over their identity. "They should always be able to refer to it, update it, or even hide it" \cite{criteria-10}.
    \item \textbf{Access.} \textit{Users must have access to their own data.} \\
    All personal claims and data should be easily retrievable for a user. No personal data is hidden for the user.
    \item \textbf{Transparency.} \textit{Systems and algorithms must be transparent.} \\
    SSI solutions and their algorithms should be open in how they function, how they are managed and updated. "The algorithms should be free, open-source, well-known, and as independent as possible of any particular architecture" \cite{criteria-10}.
    \item \textbf{Persistence.} \textit{Identities must be long-lived.} \\
    Identities can only be removed by the user. Claims can be updated and removed, but the identity that belongs to these claims should be long-lived.
    \item \textbf{Portability.} \textit{Information and services about identity must be transportable.} \\
    An identity should not be held solely by a third party. It should be transportable, since third parties may disappear.
    \item \textbf{Interoperability.} \textit{Identities should be as widely usable as possible.} \\
    A true SSI is globally usable and not limited to certain niches. 
    \item \textbf{Consent.} \textit{Users must agree to the use of their identity.} \\
    Claims and data cannot be shared without the user's consent. Users are in control of the sharing of their data.
    \item \textbf{Minimalization.} \textit{Disclosure of claims must be minimized.} \\
    Only the necessary data must be shared, when sharing some part of an identity. For example only sharing being older than 18, instead of your date of birth. This property fits well with zero-knowledge proofs.
    \item \textbf{Protection.} \textit{The rights of users must be protected.}\\
    The rights and freedoms of individuals have priority over the needs of the network.
    \item \textbf{Provable.} \textit{Claims must be shown to hold true.}\\
    It should be possible for claims to be verified, for example by trusted third parties.
\end{enumerate}

Blockchain technology already encompasses some of the eleven properties mentioned above. This makes it interesting to investigate the differences each \blockchain\ implementation has on top of these generic properties. As Q. Stokkink and J. Pouwelse have mentioned in their paper \cite{criteria-11}; data on the \blockchain\ is not deleted, only appended. This provides a \blockchain -based solution the \property{persistence} property. They also explain that the consensus algorithm that forms the basis of a \blockchain\ gives the \property{transparency} property, since it provides a global truth that is known to at least 51\% of the network. What is not mentioned here, is the possibility of \blockchain\ as an underlying system and the functionality on top of that not being transparent. Any additional non-transparent functionality will be taken into account in our evaluation.

\section{Evaluation}
\label{sec:evaluation}
After selection of the implementations that at first hand look related to Self-Sovereign Identity solutions, 11 of them were included in the review. First a brief overview, indicating the unique aspects of the different implementations is provided. Finally, the properties satisfied by the implementations are presented in \autoref{tab:overview} and \autoref{tab:overviewnon} for the \blockchain -based and other variants respectively.

\subsection{Overview of the current implementations}
\label{sec:overviewImpl}
\subsubsection{Blockchain solutions}
\paragraph{uPort}
uPort is an open-source Ethereum based \blockchain\ solution \cite{uport}. An identity is created through an app on the user's phone. This app holds all the data linked to the identity. It also holds the private keys, used to sign attestations and share them with others. Storing the data locally provides \property{control} over and \property{access} to the identity for the user and enforces them to provide \property{consent} before others can use the information.

uPort identities are capable of signing attestations about other identities and publish them on the \blockchain, increasing the trustworthiness of such claims. As only other uPort-identities are able to attest, uPort lacks \property{portability}. The only other information stored publicly on the \blockchain\ is a Decentralized ID (DID). DIDs are used as the identity and link to the public key of the user. 
All the requests and responses are transferred using industry standards, representing all different kinds of private information in a structured manner providing \property{interoperability}.

The identity consists also of several smart contracts placed on the \blockchain . One is used to allow a \property{persistent} digital identity. As the private key is stored on the smart phone, losing such a device would result in losing the identity. A social system of trustees is able to vote to replace the private key, allowing the recovery of an identity.

\paragraph{IDchainZ}
IDchainZ is a proof of concept and an extension of the ChainZy Smart Ledger \cite{morris_2019}. The program enables users to hold a ‘key ring’ of certified identity documents and allows multiple external parties to add, certify and exchange know-your-customer (KYC) and anti-money laundering (AML) documentation. This is later extended to be used for all kinds of different documents.\\ 
A user grants permission to a third party to access their data. This can be limited by the user in duration, information shared and number of accesses. This allows the user to have \property{control} over it's data. The user can also revoke access to its data anytime but it will need the private keys to do this. Otherwise the data will remain in the blockchain.\\
In the paper, nothing is mentioned about the \property{minimalization} of data. 
IDchainz is however widely available since it can be accessed within a browser.

\paragraph{EverID}
EverID is a user-centric, \selfsovereign\ and value transfer solution based on \blockchain\ technology and the cryptographic underpinnings of that system \cite{everID}. In contrast to uPort, EverID's decentralized system is used to store and confirm user identity data, documentation and bio metrics. EverID facilitates verification of users by multiple third-parties and allows the secure transfer of value between members of the network. This means that claims made by users are \property{provable}. The decentralized architecture of the platform also provides personal data ownership which can only be \property{accessed} by the user. The individual's data is recorded in a manner that allows the individual \property{control} how it is shared, with whom and for how long (\property{persistence}). This sharing mechanism is enforced by smart contracts per transaction with automated resolutions. The EverID infrastructure is operated on a series of supernodes in the network. These supernodes are the host of the \blockchain. They also host the Bridge Service to allow individuals to transfer their data to an EverID app and the API Server to enable transactions from SDK-enabled devices. This makes EverID  \property{portable}. What distinguishes EverID from other solutions is that one does not need to have a device since the digital identity (combination of biometrics, government ID and third party confirmations) can be stored in the cloud. 

Nonetheless EverID does not meet the \property{minimalization} property. When data is required to verify a claim, the user will fully disclose this particular data. If one is for example interested to know whether the user is over 18, the user can choose to disclose it's complete date of birth or to not disclose it at all. EverID is also not open source so the claims made in the whitepapers are not provable. Although EverID is a blockchain-based solution we conclude they are not \property{transparant}. 

\paragraph{Sovrin}
Sovrin network is a public permissioned distributed ledger \cite{sovrin}. This means that users can see the transactions but not necessarily initiate transactions. As a result, the need for proof-of-work is removed. This shows that they use a distributed consensus protocol which focuses more on security and scalability. This solution meets also the requirement of \property{minimalization} by enabling selective disclosure of claims using zero-knowledge proofs. Therefore no data will be shared without the user's \property{consent} and this gives the identity owners \property{control} over their digital identity.

However, there are some problems: they do not seem to provide or require any (verifiable) guarantees with respect to the correct functioning of agents in the network \cite{CASovrin}. Therefore we conclude that they do not meet the 11th property \property{provable}. 
Finally Sovrin is not \property{portable} since identities are held solely by one third party, the foundation.  

\paragraph{LifeID}
LifeID is a self-sovereign digital identity platform which enable users to create their own \property{independent} online identity \cite{LifeID_2018}. Users \property{control} all the online real-world transactions where authenticating identification is needed, without the need for third-party corporations or government agencies. LifeID is used in combination with a bio-metric-capable smart phone and app. Only the user can approve third party's request for information - always needs the user's \property{consent}. LifeID utilizes zero-knowledge proofs. The data is stored on user's device and only the needed information is released when identity needs to be verified. The use of zero-knowledge proofs is maximized making LifeID meet the \property{minimalization} requirement. LifeID identity can be backed up and recovered through three different options: back up on cold storage, with close family or friends and with a trusted organization. This gives the users ability to fight against theft by being able to deactivate identity temporarily and recover through the three above mentioned options. This is also how LifeID delivers \property{protection} against identity theft to its users, by letting the authentic owner deactivate the identity and recover it at a later moment.

\paragraph{SelfKey}
SelfKey is a Self-Sovereign digital identity network \cite{SelfKey_2017}. With SelfKey, user's data is stored on a device under the owner's \property{control}. This gives the user total control over his/her own \property{independent} identity. When a third party wants to collect specific data, which is stored on the blockchain, the user can choose to reveal it. The process of doing so is similar to authentication via "linking" a Facebook account. When approving third party data collection, SelfKey makes sure that only the minimum necessary amount of data can be collected by using zero knowledge proofs. SelfKey thus fulfills the \property{consent} and \property{minimalization} properties. To authenticate an identity, Selfkey makes use of force-resilient, censorship resistant algorithms. These independent algorithms are decentralized. Identity claims by the user can be verified only by trusted entities making sure the \property{provable} property to be fulfilled.

\paragraph{Shocard}
The ShoCard Identity Management Platform is implemented using \blockchain\ \cite{shocard}. The platform consists of three main functions: Authentication of an individual or an entity, exchange of authorization, and exchange of attestation.
Shocard has the \property{existence} property because the identity of the user is first obtained with a phone number, an official document like a passport or some biometric information.\\
The user has \property{control} over its own data. It can be stored locally and only the verification of the documents is stored publicly. This leads to having the property \property{portability}. It also \property{minimizes} the data to verify an attribute of the identity. Only the user can choose to share the data. This means that \property{access} is a property of Shocard as well.\\
The algorithms of Shocard are well documented, open-source and independent on different types of \blockchain so \property{transparency} is present. Because Shocard is able to work on different blockchains, the identity of the person can remain valid even if some \blockchain\ stops to work. That is why the platform is \property{persistent} as well.
The remaining property \property{protection} does also count in the Shocard algorithm because it uses a distributed network; \blockchain.

\paragraph{Sora}
The Sora identity solution \cite{sora} is a \blockchain\ solution based upon JSON-LD standardized key-value pairs. Storage of properties per attribute enables selective disclosure of information. A mobile application is available where a user is capable of creating an arbitrary number of identities, providing pseudonymity. 

The private key is not solely stored on the mobile device but constructed from a master password. Once an identity is established it is easy to use that identity on other devices as the created key-value pairs are encrypted and stored on an \property{accessible} centralized server. Under normal circumstances the identity is stored on the mobile device, but extra \property{persistence} is provided by the backup.

A user is in full \property{control} of and always has \property{access} to its data as no one is capable of deciphering the stored key-value pairs without the private key, i.e. \property{consent}. This also \property{protects} the user from unauthorized access. The paper states that \textit{"any user with a DID is able to issue a verifiable claim about themselves or other users"} and provides the structure in which the \property{verifiable} claims should be stored.

\subsubsection{Non-Blockchain variants}
\paragraph{PDS}
Personal Data Storages (PDS) are environments where the user has full control on the access of other parties. This architecture allows for both local \cite{openPDS} and (distributed) online storage \cite{PDS}. As the data is stored locally, the query is processed on the PDS itself. Only the answer of the query is sent back, enabling complete \property{control} about the data that is used and \property{minimalisation} of the information exposed.

When the data is stored online, nodes with different tasks communicate together to \property{protect} against unauthorized access. The data is split up in undecipherable chunks and distributed over several storage nodes. Another type of node, the index node, keeps track of the mapping of the key used to decipher the information and the key of each individual chunk of private information. The audit nodes keep track of the meaning of the information, the owner of the data and with whom the information is shared.

(open)PDS does not propose standard formats to store the information in. This implies that a user has more \property{control} about what is stored for broader use-cases and provides \property{foperability}.

It is stated that the way in which the identities used by the data owners and processors are authenticated and authorized is complex and not in the scope of the paper but will be revisited at a later date. As this is not available at the moment, there is no support for the \property{provability} of the identity.

\paragraph{IRMA}
IRMA stands for I Reveal My Attributes and implements the Idemix attribute-based credential scheme \cite{8166613}. Users receive digitally signed attributes from trusted issuers like the government. This means that claims made are \property{provable}. These claims must be stored in the IRMA app on a phones (or tablet), after which the user can selectively disclose attributes to others.  This also means that their identity can not be used without the user's \property{consent}. This way IRMA puts users in \property{control} over their digital identity. IRMA meets the  (\property{minimalization}) property by using zero-knowledge proofs. Using the issuer's digital signature over the attributes the verifier can verify that the attributes were given to the user in the past, and that they have not been modified since. An authentication is needed in order to obtain a certain attribute from an issuer. Part of the digital signature on a certain attribute is an expiration date \cite{credentialsDesignABI}. This is necessary because an attribute like "I am not older than 18" may not be true after a while. IRMA's drawback is that losing your phone means losing your identity. All the attributes need to be collected again. This means that it is not \property{persistent} since no user will have the same device his or her whole life. The decision to make IRMA data-protective to this extent makes them \property{portable}: users can bring their phones wherever they want. IRMA is currently running a trial with Gemeente Nijmegen showing that they are \property{interoperable}.  

\paragraph{reclaimID}
reclaimID is an architecture that builds on top of a name system that potentially is blockchain-based \cite{reclaimid}. Depending on the underlying name system, different properties can be assigned to reclaimID. 

The architecture makes it possible for a user to create an identity (\property{existence}) and securely share its attributes, without having a centralized service provider. The attributes are added by the user and encrypted with attribute-based encryption (ABE). The users can add, delete and update their attributes, giving them full \property{control}. The user owns the private keys to their attributes, making sure the user has full \property{access} to all personal attributes. 
To ensure the protection of the user's rights, the identity and attributes are stored locally and are shared in a decentralized manner, avoiding control by a third party. This provides the \property{protection} property. The user can decide to which parties certain attributes are revealed, providing the \property{consent} property. These parties can then access these attributes, even if the user is offline (after being granted access once). This offline availability is only possible when the requesting party uses the name system. A requesting party can get access to an attribute "... through the name system or via an out-of-band exchange, for example using a web-based authorization protocol." \cite{reclaimid}. This means that a requesting party is only bound by the name system in case of offline authorization, thus otherwise being \property{portable} as the out-of-band exchange can be anything.
reclaimID is free and the source code is available online. Therefore the architecture itself is \property{transparent}, but the underlying name system may not be. The attributes are user-defined, so reclaimID can be used in any area where identity is needed (\property{interoperability}).
Initially, the claims in reclaimID are self-attested. It is possible however, to use reclaimID to share third party attested attributes. Some claims in reclaimID are thus not \property{provable}, but might be in the future since reclaimID is working on this. \property{Minimalization} is not built in the architecture, but could be implemented additionally. Finally, the \property{persistence} of the identity within reclaimID is fully dependent on the underlying name system. If a blockchain implementation is used, it would be persistent, but other implementations may lack this property.

\begin{table*}[]
\scriptsize
\caption{Table containing the reviewed blockchain-based implementations annotated with which criteria are satisfied (1) by them, and which are not (0).}
\label{tab:overview}
\begin{tabular}{@{}lccccccccccc@{}}
\toprule
\textbf{}          & \textbf{Existence} & \textbf{Control} & \textbf{Access} & \textbf{Transparency} & \textbf{Persistence} & \textbf{Portability} & \textbf{Interoperability} & \textbf{Consent} & \textbf{Minimalization} & \textbf{Protection} & \textbf{Provable} \\ \midrule
\textit{IDchainz}  & 1 & 1 & 1 & 1 & 1 & 1 & 1 & 1 & 0 & 1 & 1 \\
\textit{Uport}     & 1 & 1 & 1 & 1 & 1 & 0 & 1 & 1 & 0 & 1 & 1 \\
\textit{EverID}    & 1 & 1 & 1 & 0 & 1 & 1 & 1 & 1 & 0 & 1 & 1 \\
\textit{Sovrin}    & 1 & 1 & 1 & 1 & 1 & 0 & 1 & 1 & 1 & 1 & 0 \\
\textit{LifeID}    & 1 & 1 & 1 & 1 & 1 & 1 & 1 & 1 & 1 & 1 & 1 \\
\textit{SelfKey}   & 1 & 1 & 1 & 1 & 1 & 1 & 1 & 1 & 1 & 1 & 1 \\
\textit{Shocard}   & 1 & 1 & 1 & 1 & 1 & 1 & 1 & 1 & 1 & 1 & 1 \\
\textit{Sora}      & 1 & 1 & 1 & 1 & 1 & 1 & 1 & 1 & 1 & 1 & 1 \\ \bottomrule
\end{tabular}
\end{table*}

\begin{table*}[]
\scriptsize
\caption{Table containing the reviewed non-blockchain-based implementations annotated with which criteria are satisfied (1) by them, and which are not (0).}
\label{tab:overviewnon}
\begin{tabular}{@{}lccccccccccc@{}}
\toprule
\textit{}          & \textbf{Existence} & \textbf{Control} & \textbf{Access} & \textbf{Transparency} & \textbf{Persistence} & \textbf{Portability} & \textbf{Interoperability} & \textbf{Consent} & \textbf{Minimalization} & \textbf{Protection} & \textbf{Provable} \\ \midrule
\textit{PDS}       & 1 & 1 & 1 & 1 & 1 & 0 & 1 & 1 & 1 & 1 & 0 \\
\textit{IRMA}      & 1 & 1 & 1 & 1 & 0 & 1 & 1 & 1 & 1 & 1 & 1 \\
\textit{reclaimID} & 1 & 1 & 1 & 1 & 0 & 1 & 1 & 1 & 0 & 1 & 0 \\ \bottomrule
\end{tabular}
\end{table*}

\newpage
\section{Discussion}
\label{sec:discussion}
Although a systematic search has been performed to gather all the available implementations, most of the found and in this survey included solutions were \blockchain -based. This, however, does not limit the possibility to discuss the differences between the two types and find the most important properties.

The comparison between \autoref{tab:overview} and \autoref{tab:overviewnon} clearly shows that on average the \blockchain -based solutions fulfill more properties than the others. In this section the global differences per property between the two types of implementations will be discussed.

\subsection{Existence}
With every solution users are able to create their own identity account and most importantly, bind their own identity into the created digital instance. This is usually done by having the users storing unique identifying information either offline as properties or online in their private cloud. 
In the research process, we have explicitly filtered out all identity solutions that were not Self-Sovereign. This could have caused all the solutions to acquire this property.

\subsection{Control}
Both \blockchain\ and non-\blockchain\ based solutions claim that users have full control over their own identity. Blockchain-based solutions usually have the users identity claims put on the blockchain and give the users control of them via their own private keys. Users can manage their identity and has control over it as long as they can authenticate themselves. All the non-blockchain-based solutions store the data locally on a mobile device or on a user-controlled environment, still allowing them full control over the user's identity.

\subsection{Access}
All blockchain-based solutions have this property since a blockchain is always public, anyone can use it to trace back the authenticity of a user. However, this does not mean that all the data about the user is public but it is traceable where the data comes from.
For non-blockchain solutions, additional concepts need to be implemented to still have the access property. PDS for example solves this by creating audit nodes that keep track of the owner and with whom the data is shared. Still it is more difficult to prove that there are no hidden algorithms when not using blockchain. So blockchain is definitely a better base for this property. 

\subsection{Transparency}
A transparent solution uses open-source algorithms and has a detailed documented system. For most \blockchain -based solutions a well-known \blockchain\ is used. However, this is not enough for the whole solution to be transparent. Still almost all solutions have the transparency property. 
Detailed documentation about the inner workings of the other implementations are also available.

\subsection{Persistence}
The benefit of using \blockchain\ is that the information on the \blockchain\ is distributed and saved at every user of the system. This means that data that verifies the authentications cannot get lost. So the data is persistent when using \blockchain . However, if a private key is lost it will be difficult to change or remove the data.

Of the non-\blockchain\ implementations solely (open)PDS acquired this property, as the user is responsible and in full control over the storage of the data. IRMA does not yet provide backups, although it is confirmed to be worked on. 

\subsection{Portability}
A solution is portable when an identity is not held only by one third party. Most of the blockchain-based solutions have this property because the identity is on a distributed ledger. Also the underlying blockchain technology is often well known so it's not dependent on a third party. But this depends on what kind of data is saved on the blockchain. That is why Uport does not have the portability property since it is only possible for other uPort identities to attest.
Not only blockchain solutions have this property as IRMA and reclaimID also met this requirement. ReclaimID does this with out-of-band exchange of information, thus the information can be communicated through another medium.

\subsection{Interoperability}
Every solution discussed in this paper has the interoperability property. This is probably because each implementation is made to be widely available. The solutions that are not, are used only in niches and therefore hard to find and not used in this paper. 

\subsection{Consent}
One of the most important properties of an SSI solution is that the user must always give consent to revealing any information to third parties. This is usually achieved by having the user sign the permission to reveal identity data with its private key. All the researched solutions fulfill this property, including the non-blockchain-based solutions. In the research process, we have explicitly filtered out all identity solutions that are not identity permission centric, which goes hand in hand with self-sovereignty. This could have caused all the solutions to have this property.

\subsection{Minimalization}
We define a solution not having the minimalization property when it does not clearly provide a way to reveal only a part of the attributes. This is the case for both various blockchain and non-blockchain based SSI solutions. The minimalization property depends on the implementation and thus is independent on whether the solutions uses the blockchain or not. In this regard, there is no difference between the two groups.

\subsection{Protection}
The protection property is met in all eleven solutions in this paper. The freedom and right of the individuals are preserved. Regardless of the usage of blockchain technology, the SSI solutions are user centered so protection of them is critical.

\subsection{Provable}
A solution fulfills the provable property if it is possible to see that the claims are verified by trusted parties. The blockchain-based solutions usually put the signed claims on the blockchain so that everyone can verify whether the signatures come from the trusted third parties. Most of the blockchain based solutions fulfill this property except Sovrin, which does not provide or require any (verifiable) guarantees. 
As for the non-blockhain based solutions, PDS did not provide any information about verification, thus does not fulfill the provable property. ReclaimID does not have the claims verification system yet. 
The verification system which is used to prove the claims is usually separately implemented from the architecture. This means that it does not make a difference whether it is a blockchain or non-blockchain based solution.

\section{Conclusion}
\label{sec:conclusion}
This paper posed the question whether \blockchain -technology is a necessity for the development of Self-Sovereign Identity solutions to solve the issues with current identity management schemes. It answers this question by systematically collecting relevant implementations, which were evaluated according to criteria introduced by C. Allen and complemented by Q. Stokkink and J. Pouwelse with provability.

Another purpose of the paper is to form a basis for future work on implementations that are most promising to serve as a digital identity solution. Identity sovereignty is mentioned for the first time in 2012 \cite{criteria-11} and is a relative new field of technology. White-papers about most of the, in this survey, reviewed solutions are published since 2017. Several mentioned to initially omit some of the desired properties to work out the proof of concept. Further research is still actively being performed, likely causing the field to change in the future.

Both \blockchain -based and other Self-Sovereign Identity solutions show to fulfill most of the evaluation criteria. The importance lies in the differences between solutions in both variants. Blockchain-based solutions definitely meet more properties on average than the others. IRMA shows that it is possible to create an \selfsovereign\ solution without \blockchain -technology. We conclude that \blockchain -technology is a good foundation to build a Self-Sovereign Identity solution, but it is not a necessity.

\appendices
\section{Research query} \label{researchquery}
The query used to systematically search for papers is as follows:

\scriptsize
\begin{lstlisting}
("Self-sovereign identity" OR "Self sovereign identity")
OR (
    ("block-chain" OR "blockchain")
    AND ("identity management")
    AND ("solution" OR "implementation" OR "review" OR "survey")
)
\end{lstlisting}

\bibliographystyle{IEEEtran}
\bibliography{IEEEabrv,references}

\end{document}